\documentstyle[12pt]{article}
\begin{document}
\baselineskip=24pt
\def\rd{{\rm d}}
\newcommand{\Lamb}{\Lambda_{b}}
\newcommand{\Lamc}{\Lambda_{c}}
\newcommand{\Lamq}{\Lambda_{Q}}
\newcommand{\omeb}{\omega_{b}^{(*)}}
\newcommand{\omec}{\omega_{c}^{(*)}}
\newcommand{\omeq}{\omega_{Q}^{(*)}}
\newcommand{\ra}{\rightarrow}
\newcommand{\dsp}{\displaystyle}
\newcommand{\nn}{\nonumber}
\newcommand{\dfr}[2]{ \displaystyle\frac{#1}{#2} }
\newcommand{\pe}{p\perp}
\renewcommand{\baselinestretch}{1.5}
\begin{titlepage}
\vspace{-20ex}
\vspace{1cm}
\begin{flushright}
\vspace{-3.0ex} 
    {\sf ADP-99-49/T385} \\
%\vspace{-2.0mm}    
%       \it{}\\
\vspace{-2.0mm}
\vspace{2.0ex}
\end{flushright}

\centerline{\Large\sf $1/m_Q$ Corrections to the Bethe-Salpeter 
Equation for $\Lambda_{Q}$ in the Diquark Picture}
\vspace{3.0ex}
\centerline{\large\sf  	
X.-H. Guo$^{1,2}$, A.W. Thomas$^{1}$  and  A.G. Williams$^{1,3}$}
\vspace{3.5ex}
\centerline{\sf $^1$ Department of Physics and Mathematical Physics,}
\centerline{\sf and Special Research Center for the Subatomic Structure of
Matter,}
\centerline{\sf University of Adelaide, SA 5005, Australia}
\centerline{\sf $^2$ Institute of High Energy Physics, Academia Sinica,
Beijing 100039, China}
\centerline{\sf $^3$ Department of Physics and SCRI,} 
\centerline{\sf Florida State University, Tallahassee, FL 32306-4052}
\centerline{\sf e-mail:  xhguo@physics.adelaide.edu.au,
athomas@physics.adelaide.edu.au,}
\centerline{\sf awilliam@physics.adelaide.edu.au}
\vspace{3ex}
\begin{center}
\begin{minipage}{5in}
\centerline{\large\sf 	Abstract}
\vspace{1.5ex}
\small {Corrections of order $1/m_Q$ ($Q=b$ or $c$) 
to the Bethe-Salpeter (B-S) 
equation for $\Lambda_{Q}$ are analyzed on the assumption that 
the heavy baryon $\Lambda_Q$ is composed of a heavy quark and a scalar, 
light diquark. It is found that in addition to the one B-S scalar function
in the limit $m_Q \ra \infty$, two more scalar functions are
needed at the order $1/m_Q$. These can be related to the B-S scalar function
in the leading order.
% With the aid of the heavy quark effective
%theory (HQET), the relationship among these scalar functions are obtained.
The six form factors for the weak transition $\Lamb \ra \Lamc$ are expressed
in terms of these wave functions and the results are consistent with 
HQET to order $1/m_Q$. Assuming the kernel for the B-S equation
in the limit $m_Q \ra \infty$ to consist of a scalar confinement term and
a one-gluon-exchange term we obtain numerical solutions for the B-S wave 
functions, and hence for the $\Lamb \ra \Lamc$ form factors to order
$1/m_Q$. Predictions are given for the differential and total decay widths 
for  $\Lamb \rightarrow \Lamc l \bar{\nu}$, and also for the 
nonleptonic decay widths for $\Lamb \ra \Lamc$ plus a pseudoscalar or vector 
meson, with QCD corrections being also included.} 

\end{minipage}
\end{center}

\vspace{0.3cm}

{\bf PACS Numbers}: 11.10.St, 12.39.Hg, 14.20.Mr, 14.20.Lq 
\end{titlepage}
\vspace{0.2in}
{\large\bf I. Introduction}
\vspace{0.2in}

Heavy flavor physics provides an important area within which
to study many important physical 
phenomena in particle physics, such as the structure and interactions inside 
heavy hadrons, the heavy hadron decay mechanism, and the plausibility of 
present nonperturbative QCD models. Heavy baryons have been studied much 
less than heavy mesons, both experimentally and 
theoretically. However, more experimental data for heavy baryons is being 
accumulated \cite{opal, ua1, cdf2, lep, cdf1, particle} 
and we expect that the experimental 
situation for them will continue to improve in the near future.
On the theoretical side, heavy quark effective theory (HQET) \cite{wise}
provides a systematic way to study physical processes involving heavy hadrons.
With the aid of HQET heavy hadron physics is simplified when $m_Q \gg
\Lambda_{QCD}$. In order to get the complete physics, HQET is usually
combined with some nonperturbative QCD models which deal with dynamics
inside heavy hadrons. 

As a formally exact equation to describe the hadronic bound state, the B-S 
equation is an effective method to deal with nonperturbative 
QCD effects. In fact,
in combination with HQET, the B-S equation has already been 
applied to the heavy meson
system \cite{dai, hussainbs, abd}. The Isgur-Wise function was calculated
\cite{dai, abd} and $1/m_Q$ corrections were also considered \cite{dai}. 
In previous work \cite{bsguo1, bsguo2, bsguo3}, we established the B-S 
equations in the heavy quark limit ($m_Q \ra \infty$)
for the heavy baryons $\Lambda_Q$ and $\omega_{Q}^{(*)}$ (where $\omega=\Xi$, 
$\Sigma$ or $\Omega$ and $Q=b$ or $c$). These were assumed to be 
composed of a heavy quark, $Q$, and a light scalar and axial-vector diquark,
respectively. We found that in the limit $m_Q \ra \infty$, the B-S equations
for these heavy baryons are greatly simplified. For example,
only one B-S scalar function is needed for$\Lambda_Q$ in this limit. By
assuming that the B-S equation's kernel consists of a scalar confinement 
term and
a one-gluon-exchange term we gave numerical solutions for the B-S wave
functions in the covariant instantaneous approximation, and consequently
applied these solutions to calculate the Isgur-Wise functions for the weak 
transitions $\Lamb \ra \Lamc$ and $\Omega_{b}^{(*)} \rightarrow 
\Omega_{c}^{(*)}$. 

In reality, the heavy quark mass is not infinite. Therefore, in order to 
give more exact phenomenological predictions we have to include $1/m_Q$
corrections, especially $1/m_c$ corrections. It is the purpose of the 
present paper to analyze the $1/m_Q$ corrections to the B-S equation for 
$\Lambda_Q$ and to 
give some phenomenological predictions for its weak decays.  
As in the previous work \cite{bsguo1, bsguo2, bsguo3, guo},
we will still assume that $\Lambda_Q$ is composed of a heavy quark and a light,
scalar diquark. In this picture, the three body system is simplified to a 
two body system.

In the framework of HQET, the eigenstate of HQET Lagrangian 
$| \Lambda_Q \rangle_{\rm HQET}$ has $0^+$ light degrees of freedom. 
This leads to only one Isgur-Wise function $\xi (\omega)$ ($\omega$ is the 
velocity transfer) for $\Lamb \ra \Lamc$ in the leading order of the $1/m_Q$ 
expansion \cite{isgur, georgi, mannel, hussain, georgi2, falk}. 
When $1/m_Q$ corrections are included, another form factor 
in HQET and an unknown flavor-independent parameter 
which is defined as the mass difference $m_{\Lambda_Q}-m_Q$ in the heavy 
quark limit are involved \cite{georgi2}. 
This provides some relations among the six
form factors for $\Lamb \ra \Lamc$ to order $1/m_Q$. Consequently, 
if one form factor is determined, the other five form factors can be obtained.

Here we extend our previous work to solve the B-S equation 
for $\Lambda_Q$ to  order $1/m_Q$, in combination with the results of  
HQET. It can be shown that two B-S scalar functions are
needed at the order $1/m_Q$, in addition to the one scalar function
in the limit $m_Q \ra \infty$. The relationship among these three scalar 
functions can be found. Therefore, our numerical
results for the B-S wave function in the order $m_Q \ra \infty$ can be
applied directly to obtain the $1/m_Q$ corrections to the form factors 
for the weak transition $\Lamb \ra \Lamc$. It can be shown that the
relations among all the six form factors for $\Lamb \ra \Lamc$ in the B-S
approach are consistent with those from HQET to order $1/m_Q$.
We also give phenomenological predictions for the differential and total 
decay widths for  $\Lamb \rightarrow \Lamc l \bar{\nu}$, and for the 
nonleptonic decay widths for $\Lamb \ra \Lamc$ plus a pseudoscalar or vector 
meson. Since the QCD corrections are comparable with the $1/m_Q$ corrections,
we also include QCD corrections in our predictions. Furthermore, we discuss
the dependence of our results on the various input parameters in our model,
and present the comparison of our results with those of other models.

The remainder of this paper is organized as follows. In Section II we
discuss the B-S equation for the heavy quark and light scalar
diquark system to order $1/m_Q$ and introduce the two B-S scalar functions
appearing at this order. We also discuss the constraint on the
form of the kernel. In Section III we express the six form factors for
$\Lamb \ra \Lamc$ in terms of the B-S wave function. The consistency
of our model with HQET is discussed. We also present numerical
solutions for these form factors. 
In Section VI we apply the solutions for the  $\Lamb \ra \Lamc$ form
factors, with QCD corrections being included, 
to the semileptonic decay $\Lamb \rightarrow \Lamc l \bar{\nu}$, 
and the nonleptonic decays $\Lamb \ra \Lamc$ plus a pseudoscalar or vector 
meson. Finally, Section VI contains a summary and discussion.

\vspace{0.2in}
{\large\bf II. The B-S equation for $\Lambda_Q$ to $1/m_Q$}
\vspace{0.2in}

Based on the picture that $\Lambda_Q$ is a bound state of a heavy quark 
and a light, scalar diquark, its B-S wave function is
defined as \cite{bsguo1}
\begin{equation}
\chi(x_1, x_2, P)=\langle 0|T \psi_Q (x_1) \varphi(x_2)|\Lambda_Q (P)\rangle,
\label{2a}
\vspace{2mm}
\end{equation}
where $\psi_Q (x_1)$ and $\varphi(x_2)$ are the field operators for the heavy
quark $Q$ and the light, scalar diquark, respectively, $P=m_{\Lambda_{Q}}v$
is the total momentum of $\Lambda_Q$ and $v$ is its velocity. Let $m_Q$ and
$m_D$  be the masses of the heavy quark and the light diquark in $\Lambda_Q$, 
$p$ be the relative momentum of the two constituents,
and define $\lambda_1=\frac{m_Q}{m_Q+m_D}, 
\lambda_2=\frac{m_D}{m_Q+m_D}$. The B-S wave function in
momentum space is defined as
\begin{equation}
\chi(x_1, x_2, P)=e^{iPX}\int \frac{\rd^4 p}{(2\pi)^4}e^{ipx}\chi_P(p),
\label{2b}
\vspace{2mm}
\end{equation}
where $X=\lambda_1 x_1+\lambda_2 x_2$ is the coordinate of the center of 
mass and $x=x_1-x_2$. The momentum of the heavy quark is
$p_1=\lambda_1 P+p$ and that of the diquark is $p_2=-\lambda_2 P+p$.
$\chi_{P}(p)$ satisfies the following B-S equation\cite{lurie}
\begin{equation}
\chi_P(p)=S_F(\lambda_1 P+p)\int \frac{\rd^4q}{(2\pi)^4}K(P,p,q)\chi_P(q)
S_D(-\lambda_2 P+p),
\label{2c}
\vspace{2mm}
\end{equation}
where $K(P,p,q)$ is the kernel, which is defined as the sum of 
all the two particle irreducible diagrams with respect to the heavy quark and
the light diquark. For convenience, in the following we use the variables 
\begin{equation}
p_l\equiv v\cdot p-\lambda_2 m_{\Lambda_Q},\;\;\; p_t\equiv p-(v\cdot p)v. 
\label{2d}
\vspace{2mm}
\end{equation}
It should be noted that $p_l$ and $p_t$ are of the order $\Lambda_{\rm QCD}$.
The mass of $\Lambda_Q$ can be written in the following form with respect to 
the $1/m_Q$ expansion (from HQET):
\begin{equation}
m_{\Lambda_Q}=m_Q+m_D+E_0+\frac{1}{m_Q}E_1+O(1/m_Q^2),
\label{2e}
\vspace{2mm}
\end{equation}
where $E_0$ and $E_1/m_Q$ are binding energies at the leading and first order
in the $1/m_Q$ expansion, respectively. 
$m_D$, $E_0$ and $E_1$ are independent of $m_Q$.  

Since we are considering $1/m_Q$ corrections to the B-S equation, we expand
the heavy quark propagator $S_F(\lambda_1 P+p)$ to order $1/m_Q$. We find
\begin{equation}
S_F=S_{0 F}+ \frac{1}{m_Q} S_{1 F},
\label{2f}
\vspace{2mm}
\end{equation}
where $S_{0 F}$ is the propagator in the limit $m_Q \ra \infty$ \cite{bsguo1}
\begin{equation}
S_{0 F}=i\frac{1+\rlap/v}{2(p_l+E_0+m_D+i\epsilon)},
\label{2g}
\vspace{2mm}
\end{equation}
and 
\begin{equation}
S_{1 F}=i\left[\frac{(-E_1 +p_t^2/2) (1+\rlap/v)}{2(p_l+E_0+m_D+i\epsilon)^2}
+\frac{\rlap/p_t}{2(p_l+E_0+m_D+i\epsilon)}-\frac{1-\rlap/v}{4}\right].
\label{2h}
\vspace{2mm}
\end{equation}

It can be shown that the light diquark propagator to $1/m_Q$ still keeps its 
form in the limit $m_Q \ra \infty$,
\begin{equation}
S_{D}=\frac{i}{p_l^2-W_p^2+i\epsilon},
\label{2i}
\vspace{2mm}
\end{equation}
where $W_{p}\equiv \sqrt{p_{t}^{2}+m_{D}^{2}}$.

Similarly to Eq.(\ref{2f}), we write $\chi_P (p)$ and $K(P,p,q)$ in the
following form (to order $1/m_Q$):
\begin{equation}
\chi_P (p)=\chi_{0 P} (p)+ \frac{1}{m_Q}\chi_{1 P} (p),\;\;\;\;
K(P,p,q)=K_0 (P,p,q)+\frac{1}{m_Q}K_1 (P,p,q),
\label{2j}
\vspace{2mm}
\end{equation}
where $\chi_{1 P} (p)$ and $K_1 (P,p,q)$ arise from $1/m_Q$ corrections.
As in our previous work, we assume the kernel contains a scalar confinement 
term and a one-gluon-exchange term. Hence we have
\begin{eqnarray}
-iK_0&=&I\otimes I V_1 +v_{\mu} \otimes (p_2+p'_2)^{\mu} V_2, \nn\\
-iK_1&=&I\otimes I V_3 +\gamma_{\mu} \otimes (p_2+p'_2)^{\mu} V_4, 
\label{2k}
\vspace{2mm}
\end{eqnarray}
where $v_{\mu}$ in $K_0$ appears because of the heavy quark symmetry. 

Substituting Eqs.(\ref{2f}) and (\ref{2j}) into the B-S equation (\ref{2c})
we have the integral equations for $\chi_{0 P} (p)$ and $\chi_{1 P} (p)$
\begin{equation}
\chi_{0 P}(p)=S_{0 F}(\lambda_1 P+p)\int \frac{\rd^4q}{(2\pi)^4}K_0(P,p,q)
\chi_{0 P}(q)S_D(-\lambda_2 P+p),
\label{2l}
\vspace{2mm}
\end{equation}
and 
\begin{eqnarray}
\chi_{1 P}(p)&=&S_{0 F}(\lambda_1 P+p)\int \frac{\rd^4q}{(2\pi)^4}K_1(P,p,q)
\chi_{0 P}(q)S_D(-\lambda_2 P+p) \nn\\
&&+S_{1 F}(\lambda_1 P+p)\int \frac{\rd^4q}
{(2\pi)^4}K_0(P,p,q)\chi_{0 P}(q)S_D(-\lambda_2 P+p)\nn\\
&&+S_{0 F}(\lambda_1 P+p)
\int \frac{\rd^4q}{(2\pi)^4}K_0(P,p,q)\chi_{1 P}(q)S_D(-\lambda_2 P+p).
\label{2m}
\vspace{2mm}
\end{eqnarray}
Eq.(\ref{2l}) is what we obtained in the limit $m_Q \ra \infty$, which
together with Eq.(\ref{2g}) gives
\begin{equation}
\rlap/v \chi_{0 P}(p)=\chi_{0 P}(p),
\label{2n}
\vspace{2mm}
\end{equation}
since $\rlap/v \rlap/v=v^2=1$ and so $\rlap/v S_{0 F}=S_{0 F}$.
Therefore, $S_{0 F}(\lambda_1 P+p) \gamma_{\mu} \chi_{0 P}(q)=
S_{0 F}(\lambda_1 P+p) v_{\mu} \chi_{0 P}(q)$ in the first term of 
Eq.(\ref{2m}). So to order $1/m_Q$, the Dirac matrix $\gamma_{\mu}$
from the one-gluon-exchange term in $K_1(P,p,q)$ can still be replaced
by $v_\mu$. 

We divide $\chi_{1 P}(p)$ into two parts by defining
\begin{equation}
\chi_{1 P}(p)=\chi_{1 P}^{+}(p)+\chi_{1 P}^{-}(p),\;\;\; 
\rlap/v \chi_{1 P}^{\pm}(p)=\pm \chi_{1 P}^{\pm}(p),
\label{2o}
\vspace{2mm}
\end{equation}
i.e., $\chi_{1 P}^{+}(p)\equiv \frac{1}{2}[\chi_{1 P}(p)+\rlap/v 
\chi_{1 P}(p)]$
and $\chi_{1 P}^{-}(p)\equiv \frac{1}{2}[\chi_{1 P}(p)-\rlap/v 
\chi_{1 P} (p)]$. After writing down all the possible terms for
$\chi_{0 P}(p)$ and $\chi_{1 P}^{\pm}(p)$, and considering
the constraints on them, Eqs.(\ref{2n}) and (\ref{2o}), we obtain that 
\begin{eqnarray}
\chi_{0 P}(p)&=&\phi_{0 P}(p)u_{\Lambda_Q}(v,s), \nn\\
\chi_{1 P}^{+}(p)&=&\phi_{1 P}(p)u_{\Lambda_Q}(v,s), \nn\\
\chi_{1 P}^{-}(p)&=&\phi_{2 P}(p)\rlap/p_t u_{\Lambda_Q}(v,s), 
\label{2p}
\vspace{2mm}
\end{eqnarray}
where $\phi_{0 P}(p)$, $\phi_{1 P}(p)$ and $\phi_{2 P}(p)$ are Lorentz scalar 
functions. 

Substituting Eq.(\ref{2p}) into Eqs.(\ref{2l})(\ref{2m}) and using  
Eqs.(\ref{2g})(\ref{2h})(\ref{2i}) we have
\begin{equation}
\phi_{0 P}(p)=-\frac{1}{(p_l+E_0+m_D+i\epsilon)(p_{l}^{2}-W_{p}^{2}
+i\epsilon)} \int \frac{\rd^4q}{(2\pi)^4}K_0(P,p,q)\phi_{0 P}(q),
\label{2q}
\vspace{2mm}
\end{equation}
\begin{eqnarray}
\phi_{1 P}(p)&=&-\frac{1}{(p_l+E_0+m_D+i\epsilon)(p_{l}^{2}-W_{p}^{2}
+i\epsilon)} \int \frac{\rd^4q}{(2\pi)^4}K_0(P,p,q)\phi_{1 P}(q)\nn\\
&&-\frac{1}{(p_l+E_0+m_D+i\epsilon)(p_{l}^{2}-W_{p}^{2}
+i\epsilon)}\int \frac{\rd^4q}{(2\pi)^4}
\left[K_1(P,p,q) \right.\nn\\
&&\left. +\frac{p_t^2/2-E_1}{p_l+E_0+m_D+i\epsilon}
K_0(P,p,q)\right]\phi_{0 P}(q),
\label{2r}
\vspace{2mm}
\end{eqnarray}
and
\begin{equation}
\phi_{2 P}(p)=\frac{1}{2}\phi_{0 P}(p).
\label{2s}
\vspace{2mm}
\end{equation}

$\phi_{0 P}(p)$ is the B-S scalar function in the leading order of the $1/m_Q$
expansion, which was calculated in \cite{bsguo1}. From Eq.(\ref{2s}) 
$\phi_{2 P}(p)$ can be given in terms of $\phi_{0 P}(p)$. The numerical
solutions for $\phi_{0 P}(p)$ and $\phi_{1 P}(p)$ can be obtained by
discretizing the integration region into $n$ pieces (with $n$ sufficiently 
large). In this way, the integral equations become matrix equations
and the B-S scalar functions $\phi_{0 P}(p)$ and $\phi_{1 P}(p)$ 
become $n$ dimensional vectors. Thus $\phi_{0 P}(p)$ is the solution of
the eigenvalue equation $(A-I) \phi_{0}=0$, where $A$ is an $n \times n$
matrix corresponding to the right hand side of Eq.(\ref{2q}). In order
to have a unique solution for the ground state, the rank of $(A-I)$ should 
be $n-1$. From Eq.(\ref{2r}), $\phi_{1 P}(p)$ is the solution of 
$(A-I) \phi_{1}=B$, where $B$ is an $n$ dimensional vector corresponding to
the second integral term on the right hand side of Eq.(\ref{2r}).
In order to have solutions for $\phi_{1 P}(p)$, the rank of the augmented
matrix
$(A-I, B)$ should be equal to that of $(A-I)$, i.e., $B$ can be expressed as
linear combination of the $n-1$ linearly independent columns in $(A-I)$.
This is difficult to guarantee if $B \neq 0$, since the way to divide 
$(A-I)$ into $n$ columns is arbitrary. Therefore, we demand the following
condition in order to have solutions for $\phi_{1 P}(p)$
\begin{equation}
\int \frac{\rd^4q}{(2\pi)^4}\left[K_1(P,p,q) +\frac{p_t^2/2-E_1}
{p_l+E_0+m_D+i\epsilon} K_0(P,p,q)\right]\phi_{0 P}(q)=0.
\label{2t}
\vspace{2mm}
\end{equation}
Eq.(\ref{2t}) provides a constraint on the form of the kernel $K_1(P,p,q)$,
in which $E_1$ is also related $K_1(P,p,q)$.
In this way, $\phi_{1 P}(p)$ satisfies the same eigenvalue equation as 
$\phi_{0 P}(p)$. Therefore, we have 
\begin{equation}
\phi_{1 P}(p)=\sigma \phi_{0 P}(p),
\label{2u}
\vspace{2mm}
\end{equation}
where $\sigma$ is a constant of proportionality, with mass dimension, 
which can be 
determined by Luke's theorem \cite{luke} at the zero-recoil point in HQET. 
We will discuss it in the next section. 

Since both $\phi_{1 P}(p)$ and $\phi_{2 P}(p)$ can be related to 
$\phi_{0 P}(p)$, we can calculate the $1/m_Q$ corrections without 
explicitly solving
the integral equations for $\phi_{1 P}(p)$ and $\phi_{2 P}(p)$.
In the previous work \cite{bsguo1} $\phi_{0 P}(p)$ 
was solved by assuming that 
$V_1$ and $V_2$ in Eq.(\ref{2k}) arise from linear confinement 
and one-gluon-exchange terms, respectively. In the covariant instantaneous 
approximation, $\tilde{V}_i \equiv V_i |_{p_l=q_l},\; i=1,2$, we find
\begin{eqnarray}
\tilde{V}_1&=&\frac{8\pi\kappa}{[(p_t-q_t)^2+\mu^2]^2}-(2\pi)^3
\delta^3  (p_t-q_t)
	\int \frac{\rd^3 k}{(2\pi)^3}\frac{8\pi\kappa}{(k^2+\mu^2)^2}, \nn \\
\tilde{V}_2&=&-\frac{16\pi}{3}
	\frac{\alpha_{s}^{({\rm eff}) 2}Q_{0}^{2}}{[(p_t-q_t)^2+\mu^2]
[(p_t-q_t)^2+Q_{0}^{2}]},
\label{2v}
\vspace{2mm}
\end{eqnarray}
where $\kappa$ and $\alpha_{s}^{({\rm eff})}$ are coupling parameters related
to scalar confinement and the one-gluon-exchange diagram,
respectively. They can be related to each 
other when we solve the eigenvalue equation for $\phi_{0 P}(p)$.
The parameter $\mu$ is introduced to avoid the infra-red divergence in 
numerical calculations, and the limit $\mu \rightarrow 0$ is taken in the end.
It should be noted that in $\tilde{V}_2$ we introduced an
effective form factor, 
$F(Q^2)=\frac{\alpha_{s}^{({\rm eff})}Q_{0}^{2}}{Q^{2}+Q_{0}^{2}}$, 
to describe the internal structure of the light diquark \cite{kroll}.

Defining $\tilde{\phi}_{0 P}(p_t)=\int \frac{\rd p_l}{2\pi} \phi_{0 P}(p)$ 
the B-S equation for $\tilde{\phi}_{0 P}(p_t)$ is \cite{bsguo1}
\begin{equation}
\tilde{\phi}_{0 P}(p_t)=-\frac{1}{2(E_0-W_p+m_D)W_{p}} 
\int \frac{\rd^3q_t}{(2\pi)^3}(\tilde{V}_1 -2W_p \tilde{V}_2)
\tilde{\phi}_{0 P}(q_t),
\label{2w}
\vspace{2mm}
\end{equation}
in the covariant instantaneous approximation. The numerical results for 
$\tilde{\phi}_{0 P}(k_t)$ can be obtained from Eq.(\ref{2w}), with the 
overall normalization constant being fixed by the normalization of the
Isgur-Wise function at the zero-recoil point \cite{bsguo1}. Furthermore, 
$\phi_{0 P}(p)$ is expressed in terms of $\tilde{\phi}_{0 P}(q_t)$:
\begin{equation}
\phi_{0 P}(p)=\frac{i}{(p_l+E_0+m_D+i\epsilon)(p_{l}^{2}-W_{p}^{2}+i\epsilon)} 
\int \frac{\rd^3q_t}{(2\pi)^3}(\tilde{V}_1 +2p_l \tilde{V}_2)
\tilde{\phi}_{0 P}(q_t).
\label{2x}
\vspace{2mm}
\end{equation}

\vspace{0.2in}
{\large\bf III. $\Lamb \ra \Lamc$ form factors to $1/m_Q$}
\vspace{0.2in}

In this section we will express the six form factors for the $\Lamb \ra \Lamc$
weak transition in terms of the B-S wave function and show the consistency 
between our model and HQET.

On the grounds of Lorentz invariance, the matrix element for 
$\Lambda_b \rightarrow \Lambda_c$ can be expressed as 
\begin{eqnarray}
\langle \Lambda_c (v')|J_\mu|\Lambda_b (v)\rangle&=&
\bar{u}_{\Lambda_c}(v')[F_1(\omega)\gamma_\mu+F_2(\omega)
v_{\mu}+F_3(\omega)v'_{\mu} \nn \\
& &-(G_1(\omega)\gamma_\mu+G_2(\omega)
v_{\mu}+G_3(\omega)v'_{\mu})\gamma_5]u_{\Lamb}(v), 
\label{3a}
\vspace{2mm}
\end{eqnarray}
where $J_\mu$ is the $V-A$ weak current, $v$ and $v'$ are the velocities of 
$\Lamb$ and $\Lamc$, respectively, and $\omega=v'\cdot v$. 

The form factors $F_i$ and $G_i$ ($i=1,2,3$) are related to each other
by the following equations, to order $1/m_Q$, when HQET is applied 
\cite{georgi2}
\begin{eqnarray}
F_1&=&G_1 \left[1+\left(\frac{1}{m_c}+\frac{1}{m_b}\right)\frac{\bar{\Lambda}}
{1+\omega}\right],\nn\\
F_2&=&G_2=-G_1 \frac{1}{m_c} \frac{\bar{\Lambda}}{1+\omega},\nn\\
F_3&=&-G_3=-G_1 \frac{1}{m_b} \frac{\bar{\Lambda}}{1+\omega},
\label{3b}
\vspace{2mm}
\end{eqnarray}
where $\bar{\Lambda}$ is an unknown parameter which is defined as the
mass difference $m_{\Lambda_Q} -m_Q$ in the limit $m_Q \ra \infty$.

On the other hand, the transition matrix element of $\Lamb \rightarrow
\Lamc$ is related to the B-S wave functions of $\Lamb$ and $\Lamc$ by
the following equation
\begin{equation}
\langle \Lamc (v')|J_\mu|\Lamb (v)\rangle =\int
\frac{\rd^4p}{(2\pi)^4} \bar{\chi}_{P'}(p')\gamma_\mu (1-\gamma_5)
\chi_{P}(p)S_{D}^{-1}(p_2),
\label{3c}
\vspace{2mm}
\end{equation}
where $P$ ($P'$) is the momentum of $\Lamb$ ($\Lamc$).
$\bar{\chi}_{P'}(p')$ is the wave function of the final state $\Lamc
(v')$ which can also be expressed in terms of the three B-S scalar functions
$\phi_{0 P}(p)$, $\phi_{1 P}(p)$ and $\phi_{2 P}(p)$ in Eq.(\ref{2p})
\begin{equation}
\bar{\chi}_{P}(p)=\bar{u}_{\Lambda_Q}(v,s)\left\{
\phi_{0 P}(p)+\frac{1}{m_Q}[\phi_{1 P}(p)+\phi_{2 P}(p)\rlap/p_t]\right\}.
\label{3d}
\vspace{2mm}
\end{equation}

Substituting Eqs.(\ref{2p}) and (\ref{3d}) into Eq.(\ref{3c}) and using the
relations in Eq.(\ref{3b}) we find the following results by comparing the
$\gamma_\mu$, $\gamma_\mu \gamma_5$, $v_\mu (1-\gamma_5)$ and  
$v'_\mu (1+\gamma_5)$ terms, respectively:
\begin{eqnarray}
G_1 \left[1+\left(\frac{1}{m_c}+\frac{1}{m_b}\right)\frac{\bar{\Lambda}}
{1+\omega}\right]&=&-i \int \frac{\rd^4 k}{(2\pi)^4}\left\{\phi_{0 P'}(k')
\phi_{0 P}(k)(k_l^2-W_k^2)\right.\nn\\
&&\left.+\frac{1}{m_c}[\phi_{1 P'}(k')-(k'_l+m_D)
\phi_{2 P'}(k')]\phi_{0 P}(k)(k_l^2-W_k^2)\right. \nn\\
&&\left.+\frac{1}{m_c}(-f_1+f_2+2m_D F)+\frac{1}{m_b}(f_1-f_2)\right.\nn\\
&&\left.+\frac{1}{m_b}\phi_{0 P'}(k')
[\phi_{1 P}(k)-(k_l+m_D)\phi_{2 P}(k)](k_l^2-W_k^2)\right\}\nn\\
&&+O(1/m_Q^2),
\label{3e}
\vspace{2mm}
\end{eqnarray}
\begin{eqnarray}
G_1 &=&-i \int \frac{\rd^4 k}{(2\pi)^4}\left\{\phi_{0 P'}(k')
\phi_{0 P}(k)(k_l^2-W_k^2)+\frac{1}{m_c}(f_1+f_2)\right.\nn\\
&&\left.+\frac{1}{m_c}[\phi_{1 P'}(k')-(k'_l+m_D)
\phi_{2 P'}(k')]\phi_{0 P}(k)(k_l^2-W_k^2)
+\frac{1}{m_b}(f_1+f_2)\right.\nn\\
&&\left.+\frac{1}{m_b}\phi_{0 P'}(k')
[\phi_{1 P}(k)-(k_l+m_D)\phi_{2 P}(k)](k_l^2-W_k^2)\right\}+O(1/m_Q^2),
\label{3f}
\vspace{2mm}
\end{eqnarray}
\begin{equation}
\frac{1}{m_c}\left[i G_1 \frac{\bar{\Lambda}}{1+\omega} +2(f_1-m_D F)\right]
=O(1/m_Q^2),
\label{3g}
\vspace{2mm}
\end{equation}
\begin{equation}
\frac{1}{m_b}\left[i G_1 \frac{\bar{\Lambda}}{1+\omega} +2f_2\right]
=O(1/m_Q^2),
\label{3h}
\vspace{2mm}
\end{equation}
where we have defined $f_1$, $f_2$ and $F$ by the following equations,
on the grounds of Lorentz invariance:
\begin{equation}
\int \frac{\rd^4 k}{(2\pi)^4}\phi_{2 P'}(k')
\phi_{0 P}(k) (k_l^2-W_k^2)= F,
\label{3i}
\vspace{2mm}
\end{equation}
\begin{equation}
\int \frac{\rd^4 k}{(2\pi)^4}\phi_{2 P'}(k')
\phi_{0 P}(k)k^\mu (k_l^2-W_k^2)= f_1 v^\mu +f_2 v'^\mu.
\label{3j}
\vspace{2mm}
\end{equation}
Eq.(\ref{3j}) leads to 
\begin{equation}
f_1+f_2=\frac{1}{1+\omega}\int \frac{\rd^4 k}{(2\pi)^4}\phi_{2 P'}(k')
\phi_{0 P}(k)(k_l^2-W_k^2)(v\cdot k + v'\cdot k).
\label{3k}
\vspace{2mm}
\end{equation}

Eqs.(\ref{3e}) and (\ref{3f}) give the expression for $G_1$ to order 
$1/m_Q$. From Eqs.(\ref{3g}) and (\ref{3h}) we can see that Eq.(\ref{3e}) is 
the same as Eq.(\ref{3f}). 
%Furthermore, as will be shown later, we can check that
%the left hand sides of Eqs.(\ref{3g})(\ref{3h}) are really of the order
%$1/m_Q^2$ from our numerical evaluations. 
Therefore, we can calculate $G_1$ to $1/m_Q$ from either of these two
equations. This indicates that our model is consistent with HQET 
to order $1/m_Q$.

Substituting Eq.(\ref{3k}) into Eq.(\ref{3f}) and using Eq.(\ref{2s}) we have
\begin{eqnarray}
G_1 &=&-i \int \frac{\rd^4 k}{(2\pi)^4}\left\{\phi_{0 P'}(k')
\phi_{0 P}(k)(k_l^2-W_k^2)\right.\nn\\
&&\left.+\frac{1}{m_c}[\phi_{1 P'}(k')-\frac{1}{2}(k'_l+m_D)
\phi_{0 P'}(k')]\phi_{0 P}(k)(k_l^2-W_k^2)\right.\nn\\
&&\left.+\frac{1}{m_b}\phi_{0 P'}(k')
[\phi_{1 P}(k)-\frac{1}{2}(k_l+m_D)\phi_{0 P}(k)]
(k_l^2-W_k^2)\right.\nn\\
&&\left.+\left(\frac{1}{m_c}+\frac{1}{m_b}\right)\frac{1}{2(1+\omega)}
\phi_{0 P'}(k')\phi_{0 P}(k)(k_l^2-W_k^2)(v\cdot k+v' \cdot k)\right\}.
\label{3l}
\vspace{2mm}
\end{eqnarray}

The first term in Eq.(\ref{3l}) gives the Isgur-Wise function which was 
calculated in our earlier work \cite{bsguo1}. In order to obtain the
$1/m_Q$ corrections, we have to fix $\phi_{1 P}(k)$. Fortunately, this
can be done by applying Luke's theorem \cite{luke}. The conservation
of vector current in the case of equal masses for the initial and final
heavy quarks leads to 
\begin{equation}
G_1 ({\omega=1})=1+O(1/m_Q^2).
\label{3m}
\vspace{2mm}
\end{equation}
Thus from Eqs.(\ref{2u}), (\ref{3f}) and (\ref{3m}) we have
\begin{equation}
\sigma=0.
\label{3n}
\vspace{2mm}
\end{equation}
Therefore, $\phi_{1 P}(k)$ does not contribute to $G_1$.
%This greatly simplifies our evaluations since we do not need the concrete 
%form for the kernel $K_1 (P,p,q)$ any more. The only requirement for    
%$K_1 (P,p,q)$ is that it should satisfy the constraint Eq.(\ref{2t}).

Now we calculate $G_1$ through Eq.(\ref{3l}).
Since in the weak transition the diquark acts as a spectator, its
momentum in the initial and final baryons should
be the same, $p_2=p'_2$. Then we can show that to order $1/m_Q$
\begin{equation}
k'_l v' +k'_t=k_l v +k_t.
\label{3o}
\vspace{2mm}
\end{equation}
From Eq.(\ref{3o}) we can obtain relations between $k'_l$, $k'_t$ and 
$k_l$, $k_t$ straightforwardly:
\begin{eqnarray}
k'_l &=& k_l\omega -k_t \sqrt{\omega^2-1}{\rm cos}\theta, \nn \\
k^{'2}_{t} &=& k_{t}^{2}+k_{t}^{2}(\omega^2-1){\rm cos}^2\theta
+k_{l}^2(\omega^2-1) -2k_l k_t \omega \sqrt{\omega^2-1} {\rm cos}\theta,
\label{3p}
\vspace{2mm}
\end{eqnarray}
where $\theta$ is defined as the angle between $k_t$ and $v'_t$.

Substituting the relation between $\phi_{0 P}(p)$ and
$\tilde{\phi}_{0 P}(p_t)$ [Eq.(\ref{2x})] into Eq.(\ref{3l}), using the B-S 
equation (\ref{2w}), and integrating the $k_l$ component by selecting the 
proper contour we have
\begin{equation}
G_1(\omega)=\xi(\omega)+\frac{1}{m_c}A_c(\omega)+\frac{1}{m_b}A_b(\omega),
\label{3q}
\vspace{2mm}
\end{equation}
where
\begin{equation}
\xi(\omega) = -\int \frac{\rd^3 k_t}{(2\pi)^3} F(\omega, k_t),
\label{3r1}
\vspace{2mm}
\end{equation}
\begin{equation}
A_c(\omega) = -\int \frac{\rd^3
k_t}{(2\pi)^3}\frac{(\omega^2-1)W_k+\omega k_t \sqrt{\omega^2-1} 
{\rm cos}\theta}{2(\omega+1)} F(\omega, k_t),
\label{3r2}
\vspace{2mm}
\end{equation}
\begin{equation}
A_b(\omega) = \int \frac{\rd^3 k_t}{(2\pi)^3}\frac{k_t \sqrt{\omega^2-1} 
{\rm cos}\theta}{2(\omega+1)} F(\omega, k_t),
\label{3r3}
\vspace{2mm}
\end{equation}
and $F(\omega, k_t)$ is defined as
\begin{eqnarray}
F(\omega, k_t) &=& \frac{\tilde{\phi}_{0 P}(k_t)}
{E_0+m_D-\omega W_k-k_t\sqrt{\omega^2-1}{\rm cos}\theta} 
\int \frac{\rd^3 r_t}{(2\pi)^3} \tilde{\phi}_{0 P'}(r_t)\nn\\
&&[\tilde{V}_1(k'_t-r_t)-2(\omega W_k
+k_t\sqrt{\omega^2-1}{\rm cos}\theta)\tilde{V}_2(k'_t-r_t)]\mid _{k_l=-W_k}.
\label{3r4}
\vspace{2mm}
\end{eqnarray}

The three dimensional integrations in Eqs.(\ref{3r1}-\ref{3r3}) can be 
reduced to one dimensional integrations by using the following identities: 
\begin{equation}
\int \frac{\rd^3 q_t}{(2\pi)^3}\frac{\rho(q_{t}^{2})}{[(p_t-q_t)^2
+\mu^2]^2}=\int \frac{q_{t}^{2}\rd q_t}{4\pi^2}
\frac{2\rho(q_{t}^{2})}{(p_{t}^{2}+q_{t}^{2}+\mu^2)^2
-4p_{t}^{2}q_{t}^{2}},
\label{3s1}
\vspace{2mm}
\end{equation}
and
\begin{equation}
\int \frac{\rd^3 q_t}{(2\pi)^3}\frac{\rho(q_{t}^{2})}{(p_t-q_t)^2
+\delta^2}=\int\frac{q_{t}^{2}\rd q_t}{4\pi^2}
\frac{\rho(q_{t}^{2})}{2|p_{t}||q_{t}|}{\rm ln}
\frac{(|p_{t}|+|q_{t}|)^{2}+\delta^2}{(|p_{t}|-|q_{t}|)^{2}+\delta^2},
\label{3s2}
\vspace{2mm}
\end{equation}
where $\rho(q_{t}^{2})$ is some arbitrary function of $q_{t}^{2}$. 

In our model we have several parameters, $\alpha_{s}^{({\rm eff})}$, $\kappa$, 
$Q_{0}^{2}$, $m_D$, $E_0$ and $E_1$. 
The parameter  $Q_{0}^{2}$ can be chosen as $3.2$GeV$^2$
from the data for the electromagnetic form factor of the proton \cite{kroll}. 
As discussed in Ref.\cite{bsguo1}, we let $\kappa$ vary in the
region between 0.02GeV$^3$ and 0.1GeV$^3$.
In HQET, the binding energies should satisfy the constraint
Eq.(\ref{2e}). Note that 
$m_D+E_0$ and $E_1$ are independent of the flavor of the
heavy quark. From the B-S equation solutions in the meson
case, it has been found that the values $m_b=5.02$GeV and $m_c=1.58$GeV give
predictions which are in good agreement with experiments
\cite{dai}. Since in the b-baryon case the $O(1/m_b^2)$ corrections are
very small, we use the following equation to discuss the relations among
$m_D$, $E_0$ and $E_1$, 
\begin{equation}
m_D+E_0+\frac{1}{m_b}E_1=0.62GeV, 
\label{3u}
\vspace{2mm}
\end{equation}
where we have used $m_{\Lambda_b}=5.64$GeV.
The parameter $m_D$ cannot be determined, although there are suggestions from 
the analysis of valence structure functions that it should be around 0.7GeV
for non-strange scalar diquarks \cite{tony}. Hence we let it vary within some 
reasonable range, 0.65GeV $\sim$ 0.75GeV. In the expansion with respect to the
heavy quark mass, we roughly expect $(\frac{1}{m_b}E_1)/E_0 \sim 
\frac{\Lambda_{\rm QCD}}{m_b}$. Therefore, $E_1$ should be of the order
$\Lambda_{\rm QCD} E_0$. In our numerical calculations, we let 
$\beta (=E_1/E_0)$ change between 0.2 and 1.0. Then for some values of $m_D$ 
and $\beta$ we can determine $E_0$. Using Eqs.(\ref{3q}-\ref{3s2}) and
Eq.(\ref{3b}) we obtain numerical results for the weak decay form factors 
$F_i$, $G_i$ ($i=1,2,3$) to order $1/m_Q$. It turns out 
that the numerical results 
are very insensitive to the value of $\beta$, so we ignore this 
dependence. We also find that the dependence of $F_i$, $G_i$ 
($i=1,2,3$) on the diquark mass $m_D$ is not strong. In Fig.1 we plot
the numerical results for $F_i$ ($i=1,2,3$) for $\kappa=0.02$GeV$^3$ and
$\kappa=0.10$GeV$^3$, respectively, with $m_D=0.7$GeV. 

\vspace{0.2in}
{\large\bf IV. Applications to $\Lamb \ra \Lamc l \bar{\nu}$ and 
$\Lamb \ra \Lamc P(V)$}
\vspace{0.2in}

With the numerical results for $F_i$, $G_i$ ($i=1,2,3$) to $1/m_Q$
obtained in Sec. III, we can predict the $\Lamb \ra \Lamc$ semileptonic and 
nonleptonic weak decay widths to order $1/m_Q$. Since the QCD corrections
to these form factors are comparable with the $1/m_Q$ effects, we will
include both of them to give phenomenological predictions.

Neubert \cite{neubert} has shown that the QCD corrections to the weak decay 
form factors can be written in the following form (up to corrections of the
order $\alpha_s \bar{\Lambda}/m_Q$):
\begin{eqnarray}
\Delta F_1&=&\xi\frac{\alpha_s (\bar{m})}{\pi} v_1, \;\;\;
\Delta G_1=\xi\frac{\alpha_s (\bar{m})}{\pi} a_1,\nn\\
\Delta F_i&=&-\xi\frac{\alpha_s (\bar{m})}{\pi} v_i,\;\;\;
\Delta G_i=-\xi\frac{\alpha_s (\bar{m})}{\pi} a_i,\;\; (i=2,3),
\label{4a}
\vspace{2mm}
\end{eqnarray}
where $v_i=v_i (\omega)$ and $a_i=a_i (\omega)$ 
$(i=1,2,3)$ are the QCD corrections calculated from
the next-to-leading order renormalization group improved perturbation 
theory. The scale $\bar{m}$ is chosen such that higher-order terms
$(\alpha_s {\rm ln}(m_b/m_c))^n$ $(n > 1)$ do not contribute. Consequently,
it is not necessary to apply a renormalization group summation as far as
only numerical evaluations are concerned. It is shown that $\bar{m}$ can
be chosen as $2m_b m_c/(m_b+m_c)\simeq 2.3$GeV. 
The detailed formulae for $v_i$ and $a_i$ can be found in \cite{neubert},
which also includes a discussion on the infra-red cutoff employed in
the calculation of the vertex corrections. As in  \cite{neubert}, we choose
this cutoff to be 200MeV which is a fictitious gluon mass. Furthermore, we
use $\Lambda_{\rm QCD}=200$MeV in our numerical calculations.  
\vspace{0.5cm}

% GNUPLOT: LaTeX picture
\setlength{\unitlength}{0.240900pt}
\ifx\plotpoint\undefined\newsavebox{\plotpoint}\fi
\sbox{\plotpoint}{\rule[-0.200pt]{0.400pt}{0.400pt}}%

\vspace{0.5cm}

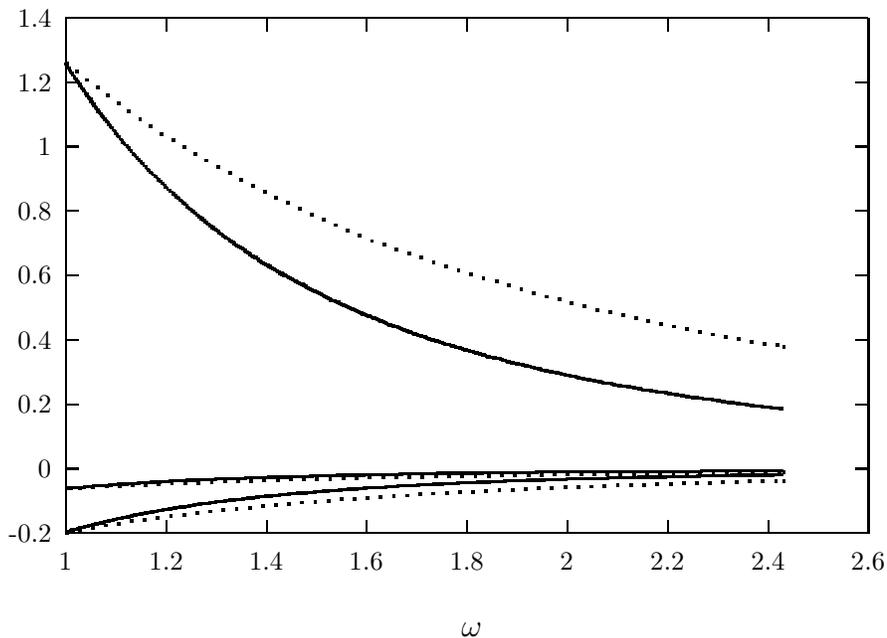
\noindent {\small Fig.1 The numerical results for $F_i$ ($i=1,2,3$) for 
$\kappa=0.02$GeV$^3$ 
(solid lines) and $\kappa=0.10$GeV$^3$ (dotted lines), with $m_D=0.7$GeV. 
From top to bottom we have $F_1$, $F_3$, and $F_2$, respectively.}

\vspace{0.2in}
{\large\bf A. Semileptonic decays $\Lamb \ra \Lamc l \bar{\nu}$}
\vspace{0.2in}

Making use of the general kinematical formulae by K\"{o}ner and Kr\"{a}mer
\cite{korner}, we find for the differential decay width of $\Lamb \ra
\Lamc l \bar{\nu}$ \cite{guo}
\begin{eqnarray}
\frac{\rd \Gamma}{\rd \omega}&=&\frac{2}{3}m_{\Lamc}^4 m_{\Lamb}A
F_1^2\sqrt{\omega^2 -1}\left\{3\omega(\eta +\eta^{-1})-2-4\omega^2
\right.\nn\\
&&\left.-\bar{\Lambda}\left(\frac{1}{m_c}+\frac{1}{m_b}\right)
[3(\eta +\eta^{-1})-4-2\omega]+\frac{\alpha_s}{\pi}v_1(\omega-1)
[3(\eta +\eta^{-1})+2-4\omega]\right.\nn\\
&&\left. +\frac{\alpha_s}{\pi}a_1(\omega+1)
[3(\eta +\eta^{-1})-2-4\omega]-\frac{\alpha_s}{\pi}(\omega^2-1)[v_2(1+\eta)
+v_3(1+\eta^{-1})\right.\nn\\
&&\left.+a_2(1-\eta)+a_3(\eta^{-1}-1)]\right\},
\label{4b}
\vspace{2mm}
\end{eqnarray}
where $\eta=m_{\Lamc}/m_{\Lamb}$ and
$A=\frac{G_{F}^2}{(2\pi)^3}|V_{cb}|^2B(\Lamc \rightarrow ab)$, with 
$|V_{cb}|$ being the Kobayashi-Maskawa matrix element. $B(\Lamc
\rightarrow ab)$ is the branching ratio for the decay $\Lamc
\rightarrow a({\frac{1}{2}}^+)+ b(0^-)$ through which $\Lamc$ is
detected, since the structure for such decays is already well known. 
It should be noted that in Eq.(\ref{4b}) $O(\alpha_s \bar{\Lambda}/m_Q)$ 
corrections have been ignored and the lepton mass is set to zero. 
The plot for $A^{-1}\frac{\rd \Gamma}{\rd
\omega}$ is shown in Fig.2 for $m_D=700$MeV, where we also show explicitly
the effects of both $1/m_Q$ and QCD corrections. For other values
of $m_D$ the results change only a little.
\vspace{0.5cm}

% GNUPLOT: LaTeX picture
\setlength{\unitlength}{0.240900pt}
\ifx\plotpoint\undefined\newsavebox{\plotpoint}\fi
\sbox{\plotpoint}{\rule[-0.200pt]{0.400pt}{0.400pt}}%

\vspace{0.5cm}

\noindent {\small Fig.2 The numerical results 
for $A^{-1}\frac{\rd \Gamma}{\rd \omega}$ 
for $\kappa=0.02$GeV$^3$ (solid lines) and $\kappa=0.10$GeV$^3$ 
(dotted lines), with $m_D=0.7$GeV. From top to bottom we have the predictions
without $1/m_Q$ and QCD corrections, with $1/m_Q$ corrections, and
with both $1/m_Q$ and QCD corrections, respectively.}

After integrating $\omega$ in eq. (\ref{4b}) we have the total decay
width for $\Lamb \rightarrow\Lamc l \bar{\nu}$. 
The numerical results are shown in Table 1 for $m_D=650$MeV, 700MeV,
750MeV and for $\kappa=0.02$GeV$^3$ (0.10GeV$^3$). $\Gamma_0$,
$\Gamma_{1/m_Q}$ and $\Gamma_{1/m_Q +{\rm QCD}}$ are the
decay widths without $1/m_Q$ and QCD corrections, with $1/m_Q$ corrections,
and with both $1/m_Q$ and QCD corrections, respectively. We have used 
$V_{cb}=0.045$ in the numerical calculations.

\begin{table}
\caption{Predictions for the decay rates for $\Lamb \rightarrow\Lamc l 
\bar{\nu}$, in units $10^{10}{\rm s}^{-1}B(\Lamc \rightarrow ab)$}
\begin{center}
\begin{tabular}{cccc}
\hline
\hline
$m_D$(GeV)&$\Gamma_0$ &$\Gamma_{1/m_Q}$ &$\Gamma_{1/m_Q +{\rm QCD}}$\\ 
\hline
0.65&4.77 (7.20)&4.26 (6.62)&3.10 (4.76) \\
\hline
0.70&5.12 (7.12)&4.60 (6.56)&3.34 (4.72) \\
\hline
0.75&5.40 (7.02)&4.89 (6.50)&3.54 (4.67) \\
\hline
\hline
\end{tabular}
\end{center}
\end{table}

We can see from Fig.2 and Table 1 that both $1/m_Q$ and QCD
corrections reduce the decay width for $\Lamb \rightarrow\Lamc l \bar{\nu}$,
and the QCD effects are even bigger. From Table 1 we can also see that the dependence of our predictions on $m_D$ is not strong.
\footnote{We note that the results 
without either $1/m_Q$ and QCD corrections in Table 1 are bigger than those 
presented in Ref.\cite{bsguo1} by about 18\%. This is because we employed 
a cutoff in the numerical integrations in Ref.\cite{bsguo1}, while the
integrations are carried out to infinity in the present work.}

\vspace{0.2in}
{\large\bf B. Nonleptonic decays $\Lamb \rightarrow \Lamc \;P \;(V)$}
\vspace{0.2in}

In this subsection we will apply the numerical solutions for the 
form factors $F_i$, $G_i$ ($i=1,2,3$) to the nonleptonic decays
$\Lamb \rightarrow \Lamc P (V)$ ($P$ and $V$
stand for pseudoscalar and vector mesons respectively). 
The Hamiltonian describing such decays reads
\begin{equation}
H_{\rm eff}=\frac{G_F}{\sqrt{2}}V_{cb}V^{*}_{UD}(a_1 O_1 +a_2 O_2),
\label{4c}
\vspace{2mm}
\end{equation}
with $O_1=(\bar D U)(\bar c b)$ and $O_2=(\bar c U)(\bar D b)$, where $U$
and $D$ are the fields for light quarks involved in the decay, and 
$(\bar q_1 q_2)=\bar q_1 \gamma_\mu (1-\gamma_5) q_2$ is understood.
The parameters $a_1$ and $a_2$ are treated as free parameters since they
involve hadronization effects. Since $\Lamb$ decays are energetic, 
the factorization assumption is applied so that one of the
currents in the Hamiltonian (\ref{4c}) is factorized out and
generates a meson\cite{bjorken, dugan}. 
Thus the decay amplitude of the two body nonleptonic decay
becomes the product of two matrix elements, one is related to the decay
constant of the factorized meson ($P$ or $V$) and the other is the weak 
transition matrix element between $\Lamb$ and $\Lamc$,
\begin{equation}
M^{{\rm fac}}(\Lambda_b \rightarrow \Lambda_c P(V))=\frac{G_F}{\sqrt{2}}
V_{cb}V^{*}_{UD}
a_1 \langle P(V)|A_\mu(V_\mu)|0\rangle \langle\Lambda_c (P')
|J^\mu|\Lambda_b (P)\rangle,
\label{4d}
\vspace{2mm}
\end{equation}
where $\langle 0|A_\mu(V_\mu)|P(V)\rangle$ are related
to the decay constants of the pseudoscalar meson  or vector meson  by
\begin{eqnarray}
\langle 0|A_\mu|P\rangle&=&if_P q_\mu, \nn\\
\langle 0|V_\mu|V\rangle&=&f_V m_V \epsilon_{\mu},
\label{4e}
\vspace{2mm}
\end{eqnarray}
where $q_\mu$ is the momentum of the meson emitted from the W-boson and 
$\epsilon_\mu$ is the polarization vector of the emitted vector meson.
It is noted that in the two-body nonleptonic weak decays
$\Lambda_b \rightarrow \Lambda_c P(V)$ there is no contribution from the
$a_2$ term since such a term corresponds to the transition of $\Lamb$ to
a light baryon instead of $\Lamc$.

On the other hand, the general form for the amplitudes of 
$\Lamb \rightarrow \Lamc P(V)$ are 
\begin{eqnarray}
M(\Lambda_b \rightarrow \Lambda_c P)&=&i\bar{u}_{\Lamc}(P')
(A+B\gamma_5) u_{\Lamb}(P), \nn \\
M(\Lambda_b \rightarrow \Lambda_c V)&=&\bar{u}_{\Lambda_c}(P')
\epsilon^{*\mu}
(A_1\gamma_\mu \gamma_5+A_2 P'_{\mu}\gamma_5+B_1\gamma_\mu
+B_2 P'_\mu)u_{\Lambda_b}(P). \nn\\
& &
\label{4f}
\vspace{2mm}
\end{eqnarray}

Alternatively, the matrix element for $\Lambda_b \rightarrow \Lambda_c$ can 
be expressed as the following on the ground of Lorentz invariance
\begin{eqnarray}
\langle\Lambda_c (P')|J_\mu|\Lambda_b (P)\rangle&=&
\bar{u}_{\Lambda_c}(P')[f_1(q^2)\gamma_\mu+if_2(q^2)\sigma_{\mu\nu}
q^\nu+f_3(q^2)q_\mu \nn \\
& &-(g_1(q^2)\gamma_\mu+ig_2(q^2)\sigma_{\mu\nu}q^\nu+g_3(q^2)q_\mu)\gamma_5]
u_{\Lambda_b}(P), 
\label{4g}
\vspace{2mm}
\end{eqnarray} 
where $f_i$, $g_i$ $(i=1,2,3)$ are the Lorentz scalars. The relations between 
$f_i, \; g_i$ and $F_i, \; G_i$ are 
\begin{eqnarray}
f_1&=&F_1+\frac{1}{2}(m_{\Lamb}+m_{\Lamc})\left( \frac{F_2}{m_{\Lamb}}
+\frac{F_3}{m_{\Lamc}} \right), \nn\\
f_2&=&\frac{1}{2}\left( \frac{F_2}{m_{\Lamb}}
+\frac{F_3}{m_{\Lamc}} \right), \nn\\
f_3&=&\frac{1}{2}\left( \frac{F_2}{m_{\Lamb}}
-\frac{F_3}{m_{\Lamc}} \right), \nn\\
g_1&=&G_1-\frac{1}{2}(m_{\Lamb}-m_{\Lamc})\left( \frac{G_2}{m_{\Lamb}}
+\frac{G_3}{m_{\Lamc}} \right), \nn\\
g_2&=&\frac{1}{2}\left( \frac{G_2}{m_{\Lamb}}
+\frac{G_3}{m_{\Lamc}} \right), \nn\\
g_3&=&\frac{1}{2}\left( \frac{G_2}{m_{\Lamb}}
-\frac{G_3}{m_{\Lamc}} \right). 
\label{4h}
\vspace{2mm}
\end{eqnarray}

The decay widths and the up-down asymmetries
for $\Lamb \rightarrow \Lamc P(V)$ are available in 
Refs.\cite{cheng}\cite{tuan}: 
\begin{eqnarray}
\Gamma(\Lamb \rightarrow \Lamc P)&=&\frac{|\vec{P}^{\prime}|}
{8\pi}\left[ \frac{(m_{\Lamb}
+m_{\Lamc})^2-m_{P}^{2}}{m_{\Lamb}^2}|A|^2+\frac{(m_{\Lamb}
-m_{\Lamc})^2-m_{P}^{2}}{m_{\Lamb}^2}|B|^2 \right], \nn\\
\alpha(\Lamb \rightarrow \Lamc P)&=&-\frac{2 |\vec{P}^{\prime}| Re(A^*B)}
{(E_{\Lamc}+m_{\Lamc})|A|^2+(E_{\Lamc}-m_{\Lamc})|B|^2},
\label{4i}
\vspace{2mm}
\end{eqnarray}
where $A$ and $B$ are related to the form factors by
\begin{eqnarray}
A&=&\frac{G_F}{\sqrt{2}}V_{cb}V^{*}_{UD}a_1 f_P [(m_{\Lamb}-m_{\Lamc}) 
f_1(m_{P}^{2})+m_P^2 f_3(m_{P}^{2})], \nn\\
B&=&\frac{G_F}{\sqrt{2}}V_{cb}V^{*}_{UD}a_1 f_P [(m_{\Lamb}+m_{\Lamc}) 
g_1(m_{P}^{2})-m_P^2 g_3(m_{P}^{2})],
\label{4j}
\vspace{2mm}
\end{eqnarray}
and
\begin{eqnarray}
\Gamma(\Lamb \rightarrow \Lamc V)&=&\frac{|\vec{P}^{\prime}|}{8\pi}
 \frac{E_{\Lamc} + m_{\Lamc}}{m_{\Lambda_{b}}}
\left[2(|{S}|^{2} + |{P_{2}}|^{2}) + \frac{E_{V}^{2}}
 {m_{V}^{2}}(|{S + D}|^{2} + |{P_{1}}|^{2}) \right], \nn\\
\alpha(\Lamb \rightarrow \Lamc V)&=&\frac{4m_{V}^{2}Re(S^{*}P_{2})
+2E_{V}^{2}Re(S+D)^{*}P_{1}}{2m_{V}^{2}(|{S}|^{2}+|{P_{2}}|^{2})
  + E_{V}^{2}(|{S+D}|^{2} + |{P_{1}}|^{2})},
\label{4k}
\vspace{2mm}
\end{eqnarray}
where
\begin{eqnarray}
 S &=& -A_{1}, \nn\\
 D &=& -\frac{|\vec{P}^{\prime}|^{2}}{E_{V}(E_{\Lamc}+m_{\Lamc})}
   (A_{1} - m_{\Lambda_{b}}A_{2}), \nn\\
 P_{1}&=& -\frac{|\vec{P}^{\prime}|}{E_{V}}(\frac{m_{\Lambda_{b}} + m_{\Lamc}}
   {E_{\Lamc} + m_{\Lamc}}B_{1} + m_{\Lambda_{b}}B_{2}), \nn\\
 P_{2} &=& \frac{|\vec{P}^{\prime}|}{E_{\Lamc}+m_{\Lamc}}B_{1},
\label{4l}
\vspace{2mm}
\end{eqnarray}
with
\begin{eqnarray}
A_1&=&-\frac{G_F}{\sqrt{2}}V_{cb}V^{*}_{UD}a_1 f_V m_V[g_1(m_{V}^{2})
+g_2(m_{V}^{2})(m_{\Lamb}-m_{\Lamc})],  \nn\\
A_2&=&-2\frac{G_F}{\sqrt{2}}V_{cb}V^{*}_{UD}a_1 f_V m_V
g_2(m_{V}^{2}), \nn\\
B_1&=&\frac{G_F}{\sqrt{2}}V_{cb}V^{*}_{UD}a_1 f_V m_V[f_1(m_{V}^{2})
-f_2(m_{V}^{2})(m_{\Lamb}+m_{\Lamc})],  \nn\\
B_2&=&2\frac{G_F}{\sqrt{2}}V_{cb}V^{*}_{UD}a_1 f_V m_V
f_2(m_{V}^{2}).
\label{4m}
\vspace{2mm}
\end{eqnarray}

Then from Eqs.(\ref{4h})-(\ref{4m}), we obtain the 
numerical results for the decay widths and asymmetry parameters. In Table 2
we list the results for $m_D=0.70$GeV. For other values of $m_D$, the results
change only a little. The numbers without (with) brackets
correspond to $\kappa=0.02$GeV$^3$ ($\kappa=0.10$GeV$^3$). Again, the
subscripts ``0'', ``$1/m_Q$'',  and ``$1/m_Q +{\rm QCD}$'' stand for
the results without $1/m_Q$ and QCD corrections, with $1/m_Q$ corrections,
and with both $1/m_Q$ and QCD corrections, respectively.
In the calculations we have taken the 
following decay constants 
$$f_\pi=132{\rm MeV}, \;\; f_K=156{\rm MeV}, \;\; f_D=200{\rm MeV}, \;\; 
f_{D_s}=241{\rm MeV},$$
$$f_\rho=216{\rm MeV}, \;\; f_{K^*}=f_\rho, \;\; f_D=f_{D^*}, \;\; 
f_{D_s}=f_{D_{s}^{*}}.$$

\begin{table}
\caption{Predictions for the decay rates (in units $10^{10}{\rm s}^{-1}
a_1^2$, which is defined in Eq.(\ref{4c})), 
and the asymmetry parameters for $\Lamb \rightarrow\Lamc P(V)$}
\begin{center}
\begin{tabular}{lcccc}
\hline
\hline
Process&$\Gamma_0$ &$\Gamma_{1/m_Q}$ &$\Gamma_{1/m_Q +{\rm QCD}}$&
$\alpha_{1/m_Q +{\rm QCD}}$\\ 
\hline
$\Lamb^0 \rightarrow \Lamc^+ \pi^-$  &0.30 (0.56)  
&0.36 (0.67) &0.29 (0.55) &-1.00 \\
\hline
$\Lamb^0 \rightarrow \Lamc^+ \rho^-$  &0.44 (0.78)  
&0.51 (0.94) &0.42 (0.77) &-0.89 \\
\hline
$\Lamb^0 \rightarrow \Lamc^+ D_{s}^{-}$ &1.03 (1.57)  
&1.16 (1.81) &1.02 (1.59) &-0.98 \\ 
\hline
$\Lamb^0 \rightarrow \Lamc^+ D_{s}^{*-}$ &0.78 (1.17)  
&0.89 (1.35) &0.76 (1.15) &-0.38 \\  
\hline
$\Lamb^0 \rightarrow \Lamc^+ K^-$  &0.022 (0.039)  
&0.026 (0.048) &0.021 (0.039) &-1.00 \\
\hline
$\Lamb^0 \rightarrow \Lamc^+ K^{*-}$  &0.023 (0.041)  
&0.027 (0.049) &0.022 (0.040) &-0.85 \\
\hline
$\Lamb^0 \rightarrow \Lamc^+ D^{-}$  &0.037 (0.057)  
&0.042 (0.066) &0.036 (0.057) &-0.98 \\
\hline
$\Lamb^0 \rightarrow \Lamc^+ D^{*-}$ &0.027 (0.041)  
&0.031 (0.048) &0.026 (0.040) &-0.42 \\ 
\hline
\hline
\end{tabular}
\end{center}
\end{table}

Since the changes for the up-down asymmetries caused by $1/m_Q$ and QCD 
corrections are very small, in Table 2 we only listed 
$\alpha_{1/m_Q +{\rm QCD}}$. Furthermore, since to order 
$O(\alpha_s \bar{\Lambda}/m_Q)$
all the six form factors $F_i$, $G_i$ ($i=1,2,3$) can be expressed by one
form factor, say $F_1$, which is canceled in $\alpha$, the up-down asymmetries 
are model independent. Therefore, $\alpha$ does not depend on $\kappa$.
It can be seen from Table 2 that the predictions for the decay widths
show a strong dependence on the parameters $\kappa$ in our model. 
In the future the experimental data will be used to
fix this parameter and test our model. 

In our previous work\cite{bsguo3, guo}, 
the $\Lamb \ra \Lamc$ semileptonic and nonleptonic 
decay widths were calculated using a hadronic wave function model in the
infinite momentum frame by combining the Drell-Yan type overlap integrals
and the results from HQET to order $1/m_Q$. Comparing the results
in our present B-S model with those in Refs.\cite{bsguo3, guo}, we
find that there is overlap between these two model predictions.
The results with $\kappa=0.02$GeV$^3$ in the present model
are close to those in Refs.\cite{bsguo3, guo} if the average transverse
momentum of the heavy quark is chosen as 400MeV. 

The Cabibbo-allowed nonleptonic decay widths have also been calculated in the 
nonrelativistic quark
model approach\cite{cheng}, where the form factors are calculated at the
zero-recoil point and then extrapolated to other $\omega$ values under
the assumption of a dipole behavior. It seems that the predictions in
this model are close to those in our present work if we choose
$\kappa=0.02$GeV$^3$.

\vspace{0.2in}
{\large\bf V. Summary and discussion}
\vspace{0.2in}

In the present work, we assume that a heavy baryon $\Lambda_Q$ is composed 
of a heavy quark, $Q$, and a scalar light diquark. Based on this picture,
we analyze the $1/m_Q$ corrections to the B-S 
equation for $\Lambda_{Q}$ which was established in the limit $m_Q \ra
\infty$ in previous work \cite{bsguo1}. We find
that in addition to the one B-S scalar function
when $m_Q \ra \infty$, two more scalar functions, $\phi_{1 P}(p)$ and
$\phi_{2 P}(p)$, are needed at order $1/m_Q$. $\phi_{2 P}(p)$ is
related to $\phi_{0 P}(p)$ directly [Eq.(\ref{2s})].
Furthermore, with the aid of the reasonable constraint
on the B-S kernel at order $1/m_Q$, Eq.(\ref{2t}), and Luke's
theorem, $\phi_{1 P}(p)$  
can also be related to the B-S scalar function in the leading order.
Hence we do not need to solve  explicitly for
$\phi_{1 P}(p)$ and $\phi_{2 P}(p)$ any more.
The B-S wave function in the leading order of $1/m_Q$ expansion
was obtained numerically by assuming the kernel for the B-S equation
in the limit $m_Q \ra \infty$ to consist of a scalar confinement term and
a one-gluon-exchange term. On the other hand, all the six form factors
for $\Lamb \ra \Lamc$ are related to each other to order $1/m_Q$, as indicated
from HQET. We determine these form factors by expressing them in
terms of the B-S wave functions. We also show explicitly that the
results from our model  are consistent with HQET to order $1/m_Q$.
We also discuss the dependence of our numerical results on the various
parameters in our model. It is found that $F_i$, $G_i$ ($i=1,2,3$) are
insensitive to the binding energy, at order $1/m_Q$, and their
dependence on the diquark mass, $m_D$, is mild. However, the numerical
solutions are very sensitive to the parameter $\kappa$.

Furthermore, we apply our solutions for the weak decay form factors to 
calculate the differential and total decay widths for the semileptonic decays
$\Lamb \rightarrow \Lamc l \bar{\nu}$, and the 
nonleptonic decay widths for $\Lamb \ra \Lamc P(V)$. The QCD corrections
are also included, and found to be comparable with the $1/m_Q$
corrections. Again the numerical results for the decay widths mostly depend on 
$\kappa$. We also compare our results with other models, including the hadronic
wave function model and the norelativistic quark model, where $1/m_Q$
corrections are also included. Generally predictions from these models are
consistent with each other if we take into account the range of model
parameters. Data from the future experiments will help to fix the parameters
and allow one to test these models.  

Besides the uncertainties from the parameters in our model, higher order
corrections such as $O(1/m_Q^2)$ and $O(\alpha_s \bar{\Lambda}/m_Q)$ 
will modify our
results. However, we expect them to be small. Furthermore, we take a
phenomenologically inspired form for the kernel of the B-S equation
and use the covariant instantaneous approximation while solving the
B-S equation. All these ans\"{a}tze should be tested by the forthcoming
experiments.

\vspace{1cm}

\noindent {\bf Acknowledgment}:

This work was supported in part by the Australian Research Council and
the National Science Foundation of China.

\vspace{2cm}

\baselineskip=20pt

\vspace{1 cm}
%\newpage

\end{document}